\documentclass[journal,twoside]{IEEEtran}

\usepackage{amsmath}
\usepackage{amssymb}
\usepackage{bm}
\usepackage{graphicx}
\usepackage{cite}
\usepackage{censor}

\newcommand{\hg}{\ensuremath{\hat{\gamma}}}
\newcommand{\me}{\ensuremath{\mathrm{e}}}
\newcommand{\diff}{\ensuremath{\mathrm{d}}}

\newcommand{\hsigma}{\ensuremath{\hat{\sigma}}}

\newtheorem{prop}{Proposition}
\newtheorem{corollary}{Corollary}

\begin{document}

\title{Outage Bound for Max-Based Downlink Scheduling With Imperfect
  CSIT and Delay Constraint}

\author{Wiroonsak Santipach,~\IEEEmembership{Senior~Member,~IEEE,} Kritsada Mamat, and Chalie Charoenlarpnopparut%
\thanks{This work was supported by joint funding from the Thailand
  Commission on Higher Education, Thailand Research Fund, and
  Kasetsart University under grant number MRG5580236, and postdoctoral funding from Faculty of Engineering, Kasetsart University under grant number 59/02/EE.}%
\thanks{W. Santipach and K. Mamat are with the Department of
  Electrical Engineering; Faculty of Engineering; Kasetsart
  University, Bangkok 10900, Thailand (email: wiroonsak.s@ku.ac.th;
  mkritsada1@gmail.com).}%
\thanks{C. Charoenlarpnopparut is with the School of Information,
  Computer, and Communication Technology, Sirindhorn International
  Institute of Technology, Thammasat University, Pathum Thani 12121,
  Thailand (email: chalie@siit.tu.ac.th).}}%
\markboth{Accepted for publication in the IEEE Communications Letter, June 2016}{Santipach \MakeLowercase{\textit{et al.}}: Outage Bound for Max-Based Downlink Scheduling With Imperfect CSIT and Delay Constraint}

\maketitle

\begin{abstract}
We consider downlink max-based scheduling in which the base station
and each user are equipped with a single antenna. In each time slot,
the base station obtains channel gains of all users and selects the
user with the largest squared channel gain. Assuming that channel
state information at the transmitter (CSIT), i.e., squared channel
gain, can be inaccurate, we derive lower bounds for probability of
outage, which occurs when a required data rate is not satisfied under
a delay constraint. The bounds are tight for Rayleigh fading and show
how required rate and CSIT error affect outage performance.
\end{abstract}

\begin{IEEEkeywords}
Outage probability, delay constraint, channel state information at
transmitter (CSIT), scheduling, downlink.
\end{IEEEkeywords}

\section{Introduction}

\IEEEPARstart{I}{mperfect} channel state information at the
transmitter (CSIT) can adversely affect the performance of wireless
communication systems~\cite[see the references
  therein]{love08}. In~\cite{vakili06}, a base station is assumed to
only have a noisy estimate of user's signal-to-noise ratio (SNR) and
thus, selects the user with the largest estimated SNR to transmit for each
time slot. An average throughput with max-based scheduling is then
determined.  In~\cite{Fettweis15}, the authors consider proportional
fairness (PF) scheduling and rate adaptation when imperfect CSIT is
assumed, and analyze the outage probability that a required data rate
is not supported in a single time slot. In~\cite{Guharoy13}, outage
probability is derived for various schedulers, assuming that the
transmitter obtains delayed feedback of channel information.

In delay-sensitive applications such as media streaming, an outage
occurs if required rate for a user is not satisfied within a given
number of time slots. In this letter, we analyze the outage
probability of max-based user scheduling with a delay constraint and
{\em imperfect} CSIT. Our results differ from those in~\cite{Johannes}
in which perfect CSIT is assumed, and in~\cite{Fettweis15} in which
delay constraint was not imposed.  We derive the lower bounds on
outage probability for flat Rayleigh fading, which are shown to be
tight for moderate required rate or when the number of users is close
to or larger than that of time slots.

\section{System Model}
\label{sys_mod}

We consider a discrete-time downlink channel in which both the base
station and each of $K$ mobile users have a single antenna. We assume
that delay spread of each user's fading channel is much smaller than
symbol period.  Thus, user's signal experiences flat fading. Let $h_j$
denote a complex channel gain and $\gamma_j \triangleq |h_j|^2$ denote the
channel power for user $j$, where $1 \le j \le K$. We assume that the
mobile users are sufficiently far apart that $h_j$'s are independent
and so are $\gamma_j$'s.

The base station is assumed to have either perfect or imperfect CSIT
of each user. Imperfection might be attributed to either channel
estimation in time-division duplex (TDD) systems or channel
quantization and feeding back in frequency-division duplex (FDD)
systems. Hence, the channel gain for user $j$ can be modeled as
follows
\begin{equation}
  h_j = \hat{h}_j + w_j
\label{error}
\end{equation}
where $\hat{h}_j$ is the imperfect gain available at the base station
and $w_j$ is the corresponding CSIT error. With max-based user
selection, the base station schedules user $k^*$ with the largest
channel power to transmit in a time slot as follows
\begin{equation}
  k^* = \arg \max_{1\le j \le K} \{\hg_j \triangleq |\hat{h}_j|^2\} .
\label{kam}
\end{equation}
Since the base station transmits to only one user in each time slot,
there is no interference among users.  The achievable rate for the
selected user in the $s$th time slot is given by
\begin{equation}
  r[s] = \log_2(1 + \rho \Gamma[s])
\end{equation}
where $\Gamma[s] = \gamma_{k^*}$ and $\rho$ is the SNR.

We also assume a delay constraint for which an outage occurs if the
achievable rate of the user over $N$ consecutive slots is less than a
rate $R$.  Given that user $k$ is selected to transmit over the set of
time slots $S_{k} \subset \{1, 2, \ldots, N\}$, the outage probability
for user $k$ is given by
\begin{equation}
  P_{k,\text{out}|S_{k}} = \Pr \left\{ \left. \frac{1}{T N}\sum_{s \in S_k}
  r[s] T  < R \right| S_{k} \right\}
\label{poex0}
\end{equation}
where $T$ is the duration of a single time slot. Channel gains are
assumed to remain constant during a time slot and fade independently
in different slots. Thus, we assume independent and identically
distributed block fading for each user and duration of a slot or block
coincides with coherence time. Consequently, the outage probability
will depend only on the number of transmitted slots denoted by
$|S_{k}|$. Thus,
\begin{equation}
  P_{k,\text{out}|S_{k}} = P_{k,\text{out}|i} = \Pr \left\{
  \left. \sum_{s \in S_k} r[s] < R N \right| |S_{k}| = i \right\} .
\label{poex1}
\end{equation}
Averaged over $N$ possible slots, the outage probability for user $k$
is given by
\begin{equation}
  P_{k, \text{out}} = \sum_{i = 0}^{N} P_{k,\text{out}|i} \Pr\{|S_k| = i\}
\label{poex}
\end{equation}
where the probability of user $k$ scheduled to transmit $i$ out of
$N$ slots is binomial and is given by~\cite{Johannes}
\begin{equation}
  \Pr\{|S_k| = i \} = {N \choose i} p^i_k (1-p_k)^{N-i} .
\label{p1bi}
\end{equation}
$p_k$ denotes the probability that user $k$ is selected to transmit in
a time slot and is given by $p_k = \Pr\{\hat{\gamma}_k \ge W_k\}$
where $W_k$ is the maximum of all other channel powers available at
the base station.  Since channel powers of different users are
independent, a cumulative distribution function (cdf) of $W_k$ can be
straightforwardly obtained as follows
\begin{equation}
  F_{W_k} (x) = \prod_{j = 1, j \ne k}^K F_{\hg_j}(x)
\label{FWk}
\end{equation}
where $F_{\hg_j}(\cdot)$ is cdf for $\hg_j$. Hence,
\begin{equation}
p_k = \int_0^\infty F_{W_k} (x) f_{\hg_k}(x) \, \diff x
\label{pk}
\end{equation}
where $f_{\hg_k}(\cdot)$ is a probability density function (pdf) for
$\hg_k$.

\section{Lower Bounds on Outage Probability}

If user $k$ is not selected to transmit in any slot ($|S_k| = 0$),
$P_{k,\text{out}|0} = 1$. For $|S_k| = 1$, we have from~\eqref{poex1},
\begin{equation}
  P_{k,\text{out}|1} = \Pr \left\{r[s] < RN \right\}.
\label{i1}
\end{equation}
For $|S_k| \ge 2$, the expression for the outage probability is not
tractable. Thus, we instead derive its lower bound by applying the
union bound to~\eqref{poex1} and obtain
\begin{align}
  P_{k,\text{out}|i} \ge \left( \Pr \left\{r[s] < RN/i \right\}
  \right)^i \quad \text{for } i \ge 2.
\label{eq_Pkt}
\end{align}

Substitute~\eqref{i1}, \eqref{eq_Pkt}, and~\eqref{p1bi}
into~\eqref{poex} to obtain the lower bound on outage probability as
follows
\begin{equation}
  P_{k, \text{out}} \ge \sum_{i=0}^N {N \choose i} p^i_k (1-p_k)^{N-i} \left( \Pr
 \left\{r[s] < RN/i \right\} \right)^i 
\label{pulb}
\end{equation}
where
\begin{equation}
  \Pr \left\{r[s] < \frac{1}{i} RN \right\} = \Pr\left\{\Gamma[s] <
  \frac{1}{\rho}(2^{RN/i} - 1)\right\}.
  \label{p1ub}
\end{equation}

For user $k$ with imperfect CSIT ($w_k \ne 0$),
\begin{align}
&\Pr\left\{\Gamma[s] < \frac{2^{RN/i} - 1}{\rho}\right\} = \Pr\left\{ \left.\gamma_k <
  \frac{2^{RN/i} - 1}{\rho}\right| \hg_k \ge W_k\right\}\\%
&= \frac{1}{p_k} \int_0^\infty \Pr \left\{\left. \gamma_k < \frac{2^{RN/i} - 1}{\rho} \right| \hg_k = x \right\}F_{W_k}(x)f_{\hg_k}(x) \, \diff x
\label{Pf1}
\end{align}
where~\eqref{Pf1} is obtained by realizing that $\gamma_k$ and $W_k$
are independent. For user $k$ with perfect CSIT ($w_k = 0$), the
outage probability can be similarly obtained and was also shown
in~\cite{Johannes}.

To tighten the bound in~\eqref{pulb}, we evaluate $P_{k,\text{out}|2}$
exactly amid increased complexity. Thus, the improved lower bound is
given by
\begin{multline}
  P_{k, \text{out}} \ge (1-p_k)^N + N p_k(1-p_k)^{N-1} \Pr\{r[s] < R N\} \\
 + \frac{1}{2}N(N-1) p_k^2(1-p_k)^{N-2} P_{k,\text{out}|2}\\
 + \sum_{i=3}^N {N \choose i} p^i_k (1-p_k)^{N-i} \left( \Pr
 \left\{r[s] < \frac{1}{i} R N \right\} \right)^i .\label{pkt}
\end{multline}
The bound is tight when the contribution of the last term
in~\eqref{pkt}, which is due to the union bound, is not significant. In
other words, that is when $N$ is not much larger than $K$.

When user $k$ is selected to transmit over 2 slots, we have
\begin{equation}
  P_{k,\text{out}|2} = \sum_{i=1}^2 \Pr\{\log_2(1 + \rho \Gamma[s_i])
  < RN\}
\end{equation}
where the selected slots $s_1, s_2 \in S_k$. We first consider user
$k$ with imperfect CSIT and obtain
\begin{multline}
  P_{k,\text{out}|2} = \Pr\{(1 + \rho \gamma_{k,s_1})(1 + \rho
  \gamma_{k,s_2}) < 2^{RN} |\\ \hg_{k,s_1} \ge W_{k,s_1},\hg_{k,s_2} \ge
  W_{k,s_2} \} . \label{Pkt}
\end{multline}
 Since channel powers in different time slots are independent, the
 conditional probability in~\eqref{Pkt} becomes
\begin{multline}
  P_{k,\text{out}|2} = \frac{1}{p_k^2} \Pr\{(1 + \rho \gamma_{k,s_1})(1 + \rho
  \gamma_{k,s_2}) < 2^{RN},\\
   \hg_{k,s_1} \ge W_{k,s_1}, \hg_{k,s_2} \ge W_{k,s_2} \}.
\end{multline}
By conditioning $\hg_{k,s_1} = x_1$ and $\hg_{k,s_2} = x_2$
and weighting the probability by the densities of $\hg_{k,s_1}$ and
$\hg_{k,s_1}$, we can compute the conditional outage probability as
follows
\begin{multline}
  P_{k,\text{out}|2} = \frac{1}{p_k^2} \iint^\infty_0
  \Pr\{(1 + \rho \gamma_{k,s_1})(1 + \rho \gamma_{k,s_2}) < 2^{RN} |\\
  \hg_{k,s_1} = x_1, \hg_{k,s_2} = x_2\}\\
  \cdot F_{W_k}(x_1) f_{\hg_k}(x_1) \, \diff x_1 F_{W_k}(x_2)
  f_{\hg_k}(x_2) \, \diff x_2. \label{pk2}
 \end{multline}

For user $k$ with perfect CSIT, $P_{k,\text{out}|2}$ can be similarly
derived.

\section{Rayleigh Fading}
\label{sec_ray}

For Rayleigh fading, the channel gain of user $j$, $h_j$, is
circularly symmetric complex Gaussian (CSCG) with zero mean and
variance $\sigma^2_j$. In TDD systems, we assume that the base station
applies linear minimum mean square error (MMSE) estimation to obtain
$\hat{h}_j$ from pilot signal. Since $h_j$ is CSCG, the estimate
$\hat{h}_j$ is also CSCG. It is well known that the MMSE estimate
$\hat{h}_j$ and the error $w_j$ are uncorrelated.  Hence, $w_j$ is
zero-mean CSCG with variance $\xi^2_j$ where $0 \le \xi^2_j \le
\sigma^2_j$, while $\hat{h}_j$ is also zero-mean CSCG with variance
$\hsigma^2_j = \sigma^2_j - \xi^2_j$.

For FDD systems, we can presume that $h_j$ is quantized at mobile $j$
and then, is fed back to the base station. To achieve the rate
distortion function with Gaussian source, the same error model used in
channel estimation can be applied in FDD systems as well. 

With the above error models, distributions of the actual and imperfect
channel power for user $j$ are exponential as follows
\begin{equation}
  F_{\gamma_j}(x) = 1 -
  \me^{-\frac{x}{2\sigma^2_j}} \text{ and } F_{\hg_j}(x) = 1 -
  \me^{-\frac{x}{2\hsigma^2_j}} . \label{Fg1}
\end{equation}
The corresponding pdf's are given by
\begin{equation}
  f_{\gamma_j}(x) = \frac{1}{2\sigma^2_j} \me^{-\frac{x}{2\sigma^2_j}}
  \text{ and } f_{\hg_j}(x) = \frac{1}{2\hsigma^2_j}
  \me^{-\frac{x}{2\hsigma^2_j}} . \label{fhx}
\end{equation}

To determine the outage probability for user $K$, we first
substitute~\eqref{Fg1} into~\eqref{FWk} and expand the product to
obtain
\begin{equation}
  F_{W_K}(x) = 1 + \sum_{i = 1}^{K-1} (-1)^i \sum_{\substack{1 \le l_1 < l_2 <
    \cdots \\< l_i \le K-1}} \me^{-\frac{1}{2} \sum\limits_{m=1}^i
    \frac{1}{\hsigma^2_{l_m}} x} .
\label{FWK}
\end{equation}
By substituting~\eqref{FWK} and~\eqref{fhx} into~\eqref{pk} and
integrating, we obtain the probability of selecting user $K$ to
transmit
\begin{equation}
  p_K = 1 + \sum_{i = 1}^{K-1} (-1)^i
  \sum_{\substack{1 \le l_1 < l_2 < \cdots \\< l_i \le K-1}} \frac{1}{1 +
    \sum\limits_{m=1}^i\frac{\hsigma^2_K}{\hsigma^2_{l_m}}} .
\end{equation}

Next we determine conditional outage probability.
\begin{prop}
\label{p1}
  Outage probability for user $K$, who is selected to transmit in 1
  out of $N$ slots, is given by
  \begin{multline}
    P_{K,\text{out}|1} = 1 - \frac{1}{p_K} \me^{-\frac{2^R -
        1}{\rho(2\hsigma^2_K + \xi_K^2)}} - \frac{1}{p_K}\sum_{i =
      1}^{K-1} (-1)^i \\
     \sum_{\substack{1 \le l_1 < l_2 < \cdots \\< l_i \le K-1}}
     \frac{1}{1 + \sum\limits_{m=1}^i\frac{\hsigma^2_K}{\hsigma^2_{l_m}}}
     \exp \left\{ -\frac{2^{RN} -1}{\frac{2\rho \hsigma^2_K}{1 +
         \sum\limits_{m=1}^i\frac{\hsigma^2_K}{\hsigma^2_{l_m}}} + \rho\xi^2_K } \right\} .
\label{Kpc}
  \end{multline}
\end{prop}
The proof is shown in the appendix. If user $K$ has perfect CSIT, we
substitute $\xi_K^2 = 0$ and $\hsigma^2_K = \sigma^2_K$ in \eqref{Kpc}
to obtain the outage probability. We remark that the complexity
of~\eqref{Kpc} mainly hinges on the last summation and increases
rapidly with $K$. Also, to determine $\Pr\{r[s] < \frac{1}{i}RN\}$ in
the bounds~\eqref{pulb} and \eqref{pkt}, we can directly
apply~\eqref{Kpc} by replacing $R$ with $R/i$.

In an ideal scenario where all users except user $K$ are in similar
fading environment and incur similar CSIT error, the expression of
outage probability given in Proposition~\ref{p1} can be reduced as
follows.
\begin{corollary}
\label{p2}
With $\hsigma^2_j = \hsigma^2$, for $\forall j \ne K$, the conditional
outage probability for user $K$ in~\eqref{Kpc} is reduced to
\begin{equation}
  P_{K,\text{out}|1} = 1 - \frac{1}{p_K} \sum_{j = 0}^{K-1}
  \frac{(-1)^j}{1 + j\frac{\hsigma^2_K}{\hsigma^2}} \exp\left\{
  -\frac{2^{RN}-1}{\frac{2 \rho \hsigma^2}{1 +
      j\frac{\hsigma^2_K}{\hsigma^2}} + \rho \xi^2_K} \right\} .
\end{equation}
\end{corollary}
Next we determine $P_{K,\text{out}|2}$ when user $K$ has imperfect
CSIT ($\xi^2_K > 0$), which is given by~\eqref{pk2}. First, we
consider the conditional probability in the integrand
of~\eqref{pk2}. Similar to the proof of Proposition~\ref{p1}, we can
show that $\gamma_{K,s_1}$ given $\hg_{K,s_1} = x_1$ and
$\gamma_{K,s_2}$ given $\hg_{K,s_1} = x_2$ are independent noncentral
Chi-squared random variables with noncentrality parameters $\lambda =
\frac{2}{\xi^2_K} x_1$ and $\frac{2}{\xi^2_K} x_2$,
respectively. Thus,
\begin{multline}
  \Pr\{(1 + \rho \gamma_{K,s_1})(1 + \rho \gamma_{K,s_2}) < 2^{RN} |
  \hg_{K,s_1} = x_1, \hg_{K,s_2} = x_2\} \\ = \iint_{(1 +
    \frac{\rho}{2}\xi^2_K z_1)(1 + \frac{\rho}{2}\xi^2_K z_2) < 2^{RN}}
  f_{\chi^2}(z_1; \frac{2}{\xi^2_K} x_1) f_{\chi^2}(z_2;
  \frac{2}{\xi^2_K} x_2) \\ \cdot \diff z_1 \diff z_2 \label{z2x1}
\end{multline}
where the pdf of a noncentral Chi-squared random variable $f_{\chi^2}$
is given by~\eqref{chi}. Substitute~\eqref{z2x1} in~\eqref{pk2} to
obtain
\begin{multline}
    P_{K,\text{out}|2} = \frac{1}{p_K^2} \int^\infty_0 \int^\infty_0
    \int_0^{\frac{2}{\rho \xi^2_K}(2^{RN} - 1)}
    \int_0^{\frac{2}{\rho \xi^2_K} \left(\frac{2^{RN}}{1 + \frac{1}{2}
        \rho \xi^2_K z_1} - 1\right)}\\ \cdot f_{\chi^2}(z_2;
    \frac{2}{\xi^2_K} x_2) \, \diff z_2  f_{\chi^2}(z_1;
    \frac{2}{\xi^2_K} x_1) \, \diff z_1 \\ \cdot F_{W_K}(x_1) f_{\hg_K}(x_1)
    \, \diff x_1 F_{W_K}(x_2) f_{\hg_K}(x_2) \, \diff x_2 .
\end{multline}

For user $K$ with perfect CSIT, similar result can be obtained. We
note that some numerical method is required to evaluate the integral
in the expression of $P_{K,\text{out}|2}$. To avoid this complexity,
the looser bound~\eqref{pulb} can be employed instead.

\section{Numerical Results}
\label{num_re}

\begin{figure}[h]
\centering
\includegraphics[width=3.4in]{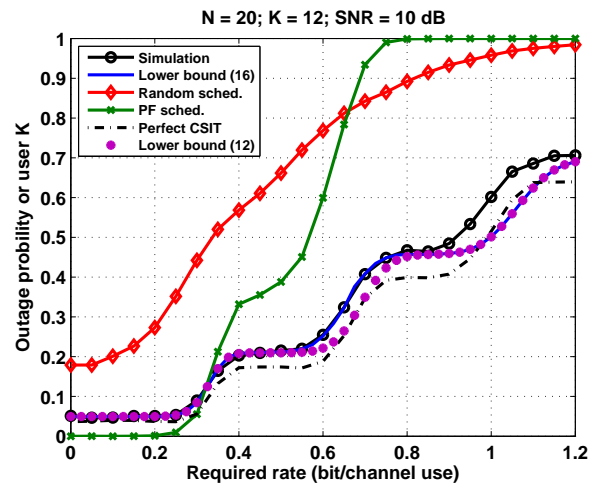}
\caption{Outage probability with different required rates.}
 \label{fig_rate}
\end{figure}

For Fig.~\ref{fig_rate}, there are 12 independent non-identically
distributed users ($K = 12$). Specifically, $\sigma^2_j = 0.9 + 0.1j$
and $\xi^2_j = 0.025(j-1)$ for $1 \le j \le 12$. Thus, user 1 is the
only user with perfect CSIT. We compare the analytical lower
bounds~\eqref{pulb} and~\eqref{pkt} with results obtained via Monte
Carlo simulations. Outage probability is shown to increase with the
required rate $R$ as expected.  We note that the bound~\eqref{pkt}
with exact evaluation of $P_{k,\text{out}|2}$ is tighter
than~\eqref{pulb} for a larger range of $R$. When rate $R$ is large,
both derived bounds are not as tight due to the union bound.

For max-based scheduling, we see that the outage performance is worse
with CSIT error. However, the outage degradation is not significant
due to relatively small error variance.  The outage performance also
displays staircase-like curves. We can attribute each step to the rate
range that certain number of transmission slots can support. For
example, the lowest rate range is supported when one or more slots are
selected and the outage probability is approximately equal to the
probability that zero slot is selected.

We also compare the results of max-based scheduling with simulation
results of random and PF scheduling~\cite{Fettweis15}. Random
scheduler performs much worse than max-based one. For PF scheduling,
the user with the largest ratio between the current rate and
cumulative rate from past slots is selected to transmit. The
performance of PF scheduler is better than that of max-based scheduler
for small $R$.

\begin{figure}
\centering
\includegraphics[width=3.4in]{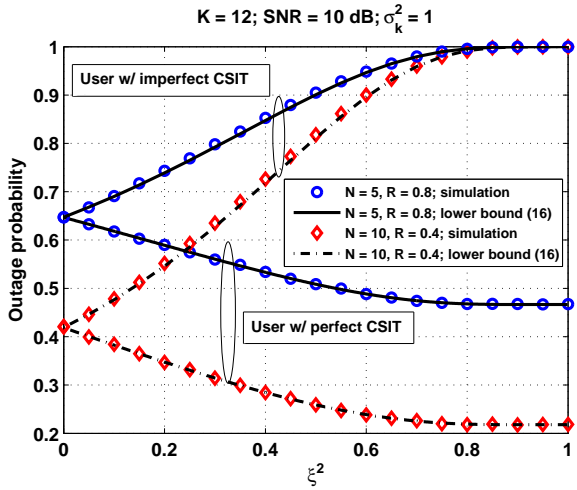}
\caption{Outage probability for user with either perfect or imperfect
  CSIT.}
 \label{fig_err}
\end{figure}

Fig.~\ref{fig_err} shows outage probability with variance of CSIT
error $\xi^2$. We assume that distributions of channel gains of all
users are identical, i.e., $\sigma^2_k = 1, \forall k$. With total 12
users ($K = 12$), 7 users have perfect CSIT ($\xi^2_k = 0$) while the
other 5 users have imperfect CSIT with the same error variance
$\xi^2$. We note that outage probability of users with CSIT error
increases with the error variance. This is due mostly to a decrease in
the probability that the user with large error will be selected to
transmit. Thus, users with perfect CSIT stand to benefit as we see a
decrease in outage.

\section{Conclusions}
\label{conclude}

The derived lower outage bounds are based on union bound and are
applicable to max-based scheduling downlink in which user channels are
independent Rayleigh fading and may not be identically distributed.
The bound is tight for small or moderate required transmission rate or
when the number of users is close to or larger than that of time
slots. The results show that when CSIT for other users is less
accurate, outage performance of user with perfect CSIT improves, and
that CSIT error can have serious impact on the outage probability.

\appendix[Proof of Proposition~\ref{p1}]

To obtain~\eqref{Kpc}, we first need to determine the conditional
probability in the integrand of~\eqref{Pf1}. Recall that $\hg_K =
|\hat{h}_K|^2 = \hat{h}_{K,r}^2 + \hat{h}_{K,i}^2 $ where
$\hat{h}_{K,r}$ and $\hat{h}_{K,i}$ are the real and imaginary parts,
respectively, of $\hat{h}_{K}$. With~\eqref{error}, the channel power
for user $K$ is given by $\gamma_K = |h_K|^2 = (\hat{h}_{K,r} +
w_{K,r})^2 + (\hat{h}_{K,i} + w_{K,i})^2$ where $w_{K,r}$ and
$w_{K,i}$ are real and imaginary parts of $w_K$ and are independent
Gaussian distributed with zero mean and variance $\frac{1}{2}
\xi^2_K$. Conditioned on $\hat{h}_{K,r}$ and $\hat{h}_{K,i}$,
$\gamma_K$ is a noncentral Chi-squared random variable with 2 degrees
of freedom. Let $\chi^2 \triangleq \frac{\gamma_K}{\frac{1}{2}
  \xi^2_K}$ be a normalized noncentral Chi-squared random variable
with a noncentrality parameter $\lambda =
\frac{2}{\xi^2_K}(\hat{h}_{K,r}^2 + \hat{h}_{K,i}^2) =
\frac{2}{\xi^2_K} x$. Thus,
\begin{align}
  \Pr \bigg\{ \gamma_K < \frac{1}{\rho}&(2^R - 1) \bigg| \hg_K =
  x\bigg\}\\
  &= 1 - \int^\infty_{\frac{2}{\xi^2_K \rho}(2^R - 1)}
  f_{\chi^2}(z; \frac{2}{\xi^2_K} x) \, \diff z \label{1iy}\\
&= 1 - Q_1 \left( \frac{1}{\xi_K} \sqrt{2 x}, \frac{1}{\xi_K}
   \sqrt{\frac{2}{\rho}(2^R - 1)} \right) . \label{1Q1}
\end{align}
where the pdf of $\chi^2$ is given by
\begin{equation}
  f_{\chi^2}(z; \lambda) = \frac{1}{2}\me^{-\frac{z + \lambda}{2}}
  I_0(\sqrt{\lambda z}), \label{chi}
\end{equation}
and $I_0(\cdot)$ denotes the zeroth-order modified Bessel function of
the first kind. The complementary cdf for $\chi^2$ in~\eqref{1iy} can
be expressed as the first-order Marcum Q-function defined
in~\cite{nuttall75} as shown in~\eqref{1Q1}. 

Finally, to obtain~\eqref{Kpc}, we substitute~\eqref{1Q1},
\eqref{FWK}, and \eqref{fhx} into~\eqref{Pf1}, and evaluate the
integrals by substituting $y = \sqrt{x}$ and applying the following
result, which can be obtained from \cite[eqs. (2) and
  (36)]{nuttall72},
\begin{equation}
  \int_0^\infty y \me^{-\frac{1}{2} p^2 y^2} Q_1(ay, b) \, \diff y =
  \frac{1}{p^2} \me^{-\frac{p^2 b^2}{2(a^2 + p^2)}}
\end{equation}
where $a$, $b$, and $p$ are constant.

 \section*{Acknowledgment}
The authors would like to thank Prof. Norbert Goertz of the
  Institute of Telecommunications, Vienna University of Technology,
  Austria, and Johannes Gonter for insightful discussion and for
  graciously hosting them during their visits.

\bibliographystyle{IEEEtran}
\bibliography{IEEEabrv,Delay}

\begin{thebibliography}{1}
\providecommand{\url}[1]{#1}
\csname url@samestyle\endcsname
\providecommand{\newblock}{\relax}
\providecommand{\bibinfo}[2]{#2}
\providecommand{\BIBentrySTDinterwordspacing}{\spaceskip=0pt\relax}
\providecommand{\BIBentryALTinterwordstretchfactor}{4}
\providecommand{\BIBentryALTinterwordspacing}{\spaceskip=\fontdimen2\font plus
\BIBentryALTinterwordstretchfactor\fontdimen3\font minus
  \fontdimen4\font\relax}
\providecommand{\BIBforeignlanguage}[2]{{%
\expandafter\ifx\csname l@#1\endcsname\relax
\typeout{** WARNING: IEEEtran.bst: No hyphenation pattern has been}%
\typeout{** loaded for the language `#1'. Using the pattern for}%
\typeout{** the default language instead.}%
\else
\language=\csname l@#1\endcsname
\fi
#2}}
\providecommand{\BIBdecl}{\relax}
\BIBdecl

\bibitem{love08}
D.~J. Love, R.~W. Heath, Jr., V.~K.~N. Lau, D.~Gesbert, B.~D. Rao, and
  M.~Andrews, ``An overview of limited feedback wireless communication
  systems,'' \emph{{IEEE} J. Sel. Areas Commun.}, vol.~26, no.~8, pp.
  1341--1365, Oct. 2008.

\bibitem{vakili06}
A.~Vakili, M.~Sharif, and B.~Hassibi, ``The effect of channel estimation error
  on the throughput of broadcast channels,'' in \emph{Proc. IEEE Int. Conf. on
  Acoustics, Speech and Signal Processing (ICASSP)}, vol.~4, Toulouse, France,
  May 2006, pp. IV--IV.

\bibitem{Fettweis15}
R.~Fritzsche, P.~Rost, and G.~P. Fettweis, ``Robust rate adaptation and
  proportional fair scheduling with imperfect {CSI},'' \emph{{IEEE} Trans.
  Wireless Commun.}, vol.~14, no.~8, pp. 4417--4427, Aug. 2015.

\bibitem{Guharoy13}
S.~Guharoy and N.~B. Mehta, ``Joint evaluation of channel feedback schemes,
  rate adaptation, and scheduling in {OFDMA} downlinks with feedback delays,''
  \emph{{IEEE} Trans. Veh. Technol.}, vol.~62, no.~4, pp. 1719--1731, May 2013.

\bibitem{Johannes}
J.~Gonter, N.~Goertz, and A.~Winkelbauer, ``Analytical outage probability for
  max-based schedulers in delay-constrained applications,'' in \emph{Proc.
  Wireless Days 2012}, Dublin, Ireland, Nov. 2012.

\bibitem{nuttall75}
A.~H. Nuttall, ``Some integrals involving the ${Q}_{M}$ function,''
  \emph{{IEEE} Trans. Inf. Theory}, vol.~21, no.~1, pp. 95--96, Jan. 1975.

\bibitem{nuttall72}
------, ``Some integrals involving the ${Q}$ function,'' Naval Underwater
  Systems Center, New London, Connecticut, USA, Technical Report 4755, Apr.
  1972.

\end{thebibliography}

\end{document}